\documentclass[9pt,onecolumn,twoside]{article}

\usepackage{lipsum}
\usepackage{authblk}
\usepackage{fancyhdr}
\usepackage{amssymb}
\usepackage{amsmath}
\usepackage{xfrac}
\usepackage[english,plain]{fancyref}
\usepackage[margin=1in]{geometry}

\title{Interferometric method for the complete characterization of highly chirped ultrabroadband pulses}

\author[1,2,*]{Adam S Wyatt}
\author[1]{Pedro Oliveira}
\author[1]{Ian O Musgrave}

\affil[1]{Central Laser Facility, STFC Rutherford Appleton Laboratory, Harwell OX11 0QX, UK}
\affil[2]{Clarendon Laboratory, Department of Physics, University of Oxford, Oxford OX1 3PU, UK}

\affil[*]{Corresponding author: adam.wyatt@stfc.ac.uk}

\begin{document}

\maketitle
\thispagestyle{fancy}

\abstract{
This article describes an interferometric method, called ``Chirped Heterodyne Interferometry for Measuring Pulses'' (CHIMP), for the complete characterization of highly (monotonically) chirped ultrabroadband optical pulses. CHIMP provides the spectrally dependent group delay dispersion (GDD) of the a chirped test pulse (CTP) via a simple direct algorithm and is verified via second harmonic generation (SHG) simulations and experimental measurements of pulses centred at 800\,nm with a bandwidth of 55\,nm stretched to 32\,ps at the 1\% intensity level, corresponding to a time-bandwidth product of 830.
}

\section{Introduction}
\label{sec:INTRO}
The complete characterization of ultrashort optical pulses is a key capability in the design, optimization and implementation of any ultrafast laser system or experiment and an accurate means to quantify the complex spectral or temporal amplitudes of the field enable maximum information to be extracted from dynamic measurements. Developments in ultrafast metrology have played a significant role in the development of ultrafast laser systems, and in increasing the information and understanding of ultrafast experiments~\cite{assion1998control,levis2001selective}.

There are a variety of applications which make use of chirped pulses. They are most commonly employed in high power short pulse [optical parametric] chirped pulse amplification ([OP]CPA) laser systems~\cite{strickland1985compression,backus1998high,ross1997prospects, cerullo2003ultrafast, butkus2004progress}. OPCPA systems have enabled the amplification of extremely large bandwidths, supporting few-cycle pulses~\cite{herrmann2009generation} but are susceptible to dispersion and pump intensity reshaping effects that can limit the compressibility of the amplified pulses~\cite{tavella2007dispersion}. Careful design and implementation of complex stretcher/compressor configurations is required in order to compensate for the material dispersion and OPA phase, hence composite systems are typically employed utilizing a combination of prism, grating and bulk stretchers/compressors in conjunction with an adaptive element~\cite{mikhailova2011ultra}. The cost and complexity of compressing these pulses makes it desirable to have a means to measure the amplified chirped pulses at various locations along the laser chain in order to diagnose and optimize each element of the chain in situations where a compressed pulse is not available and cannot easily be generated. Extremely high contrast lasers systems use picosecond duration pump pulses, and therefore place even more stringent demands on controlling the optical dispersion throughout the laser chain~\cite{klingebiel2013picosecond,skrobol2012broadband,rigaud2016supercontinuum}. 

One particular laser system of personal importance where the effects of the nonlinear chirp are especially acute is in chirp-compensated (CC-) OPCPA schemes~\cite{tang2008optical, wyatt2015ultra}. In CC-OPCPA, the instantaneous frequencies of the pump and seed are tuned to compensate the dispersion of the nonlinear medium. This then enables the amplification of large bandwidths in a collinear geometry, allowing the idler to be used without having to compensate for angular dispersion --- this is particularly advantageous for passive carrier-envelope phase (CEP) stabilization and frequency tuning of the system~\cite{vozzi2007millijoule, brida2009few, thai2011sub} and has the potential to improve the contrast in the compressed pulses. Since the frequency-dependent group delay (GD) of the pump and seed need to be matched to better than 100\,fs over the whole duration of the pulse (typically >10\,ps), a coarse measurement of the pulse duration is not sufficient and accurate measurement of the nonlinear dispersion is required.

Other applications of chirped pulses include chirp-assisted sum frequency generation (CA-SFG)~\cite{osvay1999efficient}, dispersive Fourier transform spectroscopy (DFTS)~\cite{goda2013dispersive} and telecommunications~\cite{boivin1997206}. A normal requirement for DFTS is that the pusles are linear chirped to prevent distortion of the measured data, but this constraint can be relaxed provided the chirp is monotonic known~\cite{mahjoubfar2015design}. A means to easily and accurately characterize the chirped pulses is essential to ensure the information is accurately extracted from the data.

A wide variety of complete characterization methods already exist~\cite{walmsley2009characterization}. Although it is possible to measure the electric waveform of ultrashort pulses~\cite{frank2012invited, kim2013petahertz, wyatt2016attosecond}, these methods typically require near few-cycle waveforms and strong-field phenomena and so are not applicable to highly chirped pulses. The most well known general methods include spectral phase interferometry for direct electric field reconstruction (SPIDER)~\cite{iaconis1998spectral}, frequency-resolved optical gating (FROG)~\cite{trebino1997measuring}, multiphoton intrapulse interference phase scan (MIIPS)~\cite{lozovoy2004multiphoton} and dispersion scan (DS)~\cite{miranda2012simultaneous}. Extending their capabilities to measure highly chirped pulses is a demanding task. A limiting factor in SPIDER is the requirement to frequency mix with a pulse which is quasi-monochromatic over the duration of the test pulse. DS by design measures the pulse around zero net group delay dispersion (GDD), and MIPS requires the application of a phase function that can compensate the frequency dependent group delay across the whole spectrum. 

Although FROG has been used to measure nanosecond duration optical pulses~\cite{bowlan2011complete} and more recently been applied to the measurement of highly chirped pulses~\cite{wyatt2016frequency}, it is not ideally suited to the task. For extremely large stretch factors, employing a single-shot version becomes problematic, whereas a scanning version becomes impractical and unstable. The reconstruction uses an iterative procedure and can be quite slow to converge. Additionally, the reconstructed spectral phase is susceptible to variations in the measured spectral intensity, for example due to an improper intensity calibration.

In this paper I shall describe a new self-referenced interferometric method for the characterization of monotonically chirped broadband pulses. I have named the method ``Chirped Heterodyne Interferometry for Measuring Pulses'' (CHIMP). The method is conceptually similar to SPIDER, although with several significant differences. CHIMP enables the extraction of the GDD via a simple, rapid and direct procedure based on the Takeda algorithm~\cite{takeda1982fourier} in a similar manner to SPIDER and can potentially be extended to enable single-shot acquisition.
A schematic of the CHIMP concept is shown in \fref{fig:CONCEPT}. In the particular implementation displayed, which matches the measurements performed below, three copies of the chirped test pulse (CTP) are required, but the data can be measured on a linear array spectrometer.

\section{Method}
\label{sec:METHOD}

\begin{figure}[htbp]
\centering
\fbox{\includegraphics[width=\linewidth]{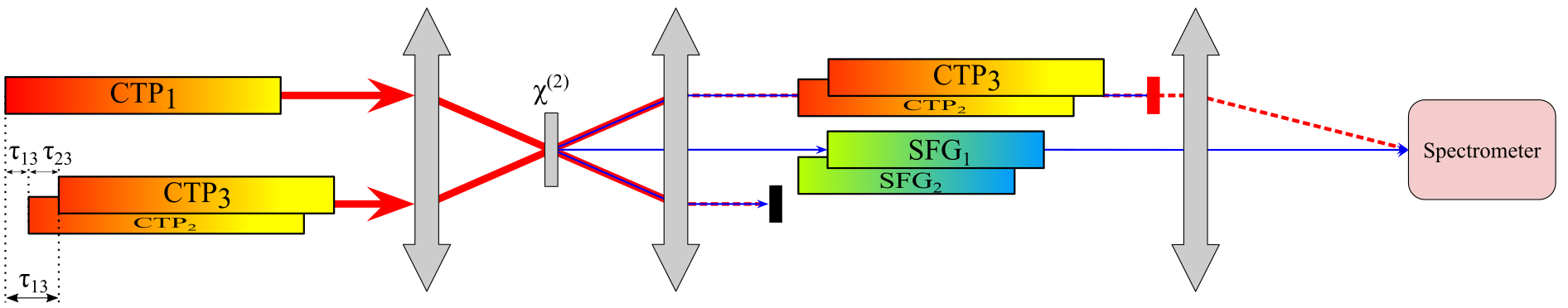}}
\caption{Schematic of CHIMP concept: two time-delayed chirped test pulses (CTP) are mixed with a third CTP in a $\chi^{(2)}$ nonlinear crystal. The two sum-frequency generation (SFG) signals are then imaged and interfered onto the entrance slit of a spectrometer. Spectral fringes allow the interferometric phase to be retrieved using the Takeda algorithm.}
\label{fig:CONCEPT}
\end{figure}

A monotonically chirped pulse is defined as one in which the group delay dispersion (GDD), $\phi^{\prime\prime}(\omega)$, is of constant sign and large magnitude relative to the bandwidth of the finest spectral feature, $\Delta\omega_\text{min}$:
\begin{equation}
    \label{eq:SPA_CRITERION}
    \left|\phi^{\prime\prime}(\omega)\right| \gg 2\pi/\Delta\omega_\text{min}^2.
\end{equation}
If this criterion is satisfied, then the instantaneous frequency is mapped to time according to $\tau(\omega) = \phi^\prime(\omega)$ and therefore the group delay of two pulse replicas delayed in time are related according to
\begin{equation}
    \label{eq:GD}
    \phi^\prime(\omega-\delta\omega) = \phi^\prime(\omega+\delta\omega) + \tau.
\end{equation}
After mixing the two time-delayed pulses in a $\chi^{(2)}$ nonlinear crystal in the zero depletion limit, the spectral phase of a the sum frequency generation (SFG) signal is given as
\begin{equation}
    \label{eq:SFG}
    \phi_\text{SFG}(2\omega,\tau) = \phi(\omega - \delta\omega) + \phi(\phi + \delta\omega) + (\omega + \delta\omega)\tau.
\end{equation}
Peforming a Taylor expansion of \fref{eq:GD} and rearranging yields $\delta\omega\simeq-\tfrac{1}{2}\tau/\phi^{\prime\prime}(\omega)$. Substituting this solution into a Taylor expansion of \fref{eq:SFG} gives an approximate form for the SFG phase as
\begin{equation}
    \label{eq:SFG_APPROX}
    \phi_\text{SFG}(2\omega,\tau) \simeq 2\phi(\omega) + \omega\tau - \frac{\tau^2}{4\phi^{\prime\prime}(\omega)}.
\end{equation}
In order to be able to measure this spectral phase, it is necessary to interfere it with another pulse as depicted in \fref{fig:CONCEPT}. The spectral phase between SFG$_{12}$ and SFG$_{13}$, which I denote as the CHIMP phase $\theta(\omega)$, can therefore be written as
\begin{align}
    \label{eq:CHIMP_PHASE}
    \theta(2\omega, \delta\tau, \tau_{12}) &= \phi_\text{SFG}(2\omega,\tau_{13}) - \phi_\text{SFG}(2\omega,\tau_{12}-\delta\tau)\\\nonumber
    &= \frac{2\tau_{12}\delta\tau - \delta\tau^2}{4\phi^{\prime\prime}(\omega)} + \omega\delta\tau
\end{align}
where $\tau_{nm}$ is the delay between CTP$_n$ and CTP$_m$, and $\delta\tau = \tau_{23} = \tau_{12}-\tau_{13}$ is the delay between the collinear CTP$_2$ and CTP$_3$. It is fairly straightforward to calibrate $\delta\tau$ simply by measuring the interference pattern of the fundamental pulses 
and extracting the linear phase term from the interferogram. Due to the non-collinear geometry, it is not possible to uniquely define the delay, $\tau_{12}$, between CTP$_1$ and CTP$_2$ since this depends on the transverse spatial position. It is possible to eliminate this term by taking the difference between two CHIMP phases with a temporal shift, $\Delta\tau$, in CTP$_1$ (i.e. varying $\tau_{12}$ whilst keeping $\delta\tau$ fixed). Rearranging \fref{eq:CHIMP_PHASE} from two such measurements gives the CTP GDD as
\begin{align}
    \label{eq:CHIMP_RECONSTRUCTION}
    \Delta\theta(2\omega, \delta\tau, \Delta\tau) &= \theta(2\omega, \delta\tau, \tau_{12}) - \theta(2\omega, \delta\tau, \tau_{12}-\Delta\tau)\\\nonumber
    \phi^{\prime\prime}(\omega) &= \frac{\Delta\tau\delta\tau}{2\Delta\theta(2\omega, \delta\tau, \Delta\tau)}.
\end{align}

\section{Simulations}
\label{sec:SIMULATIONS}

\begin{figure}[htbp]
\centering
\fbox{\includegraphics[width=\linewidth]{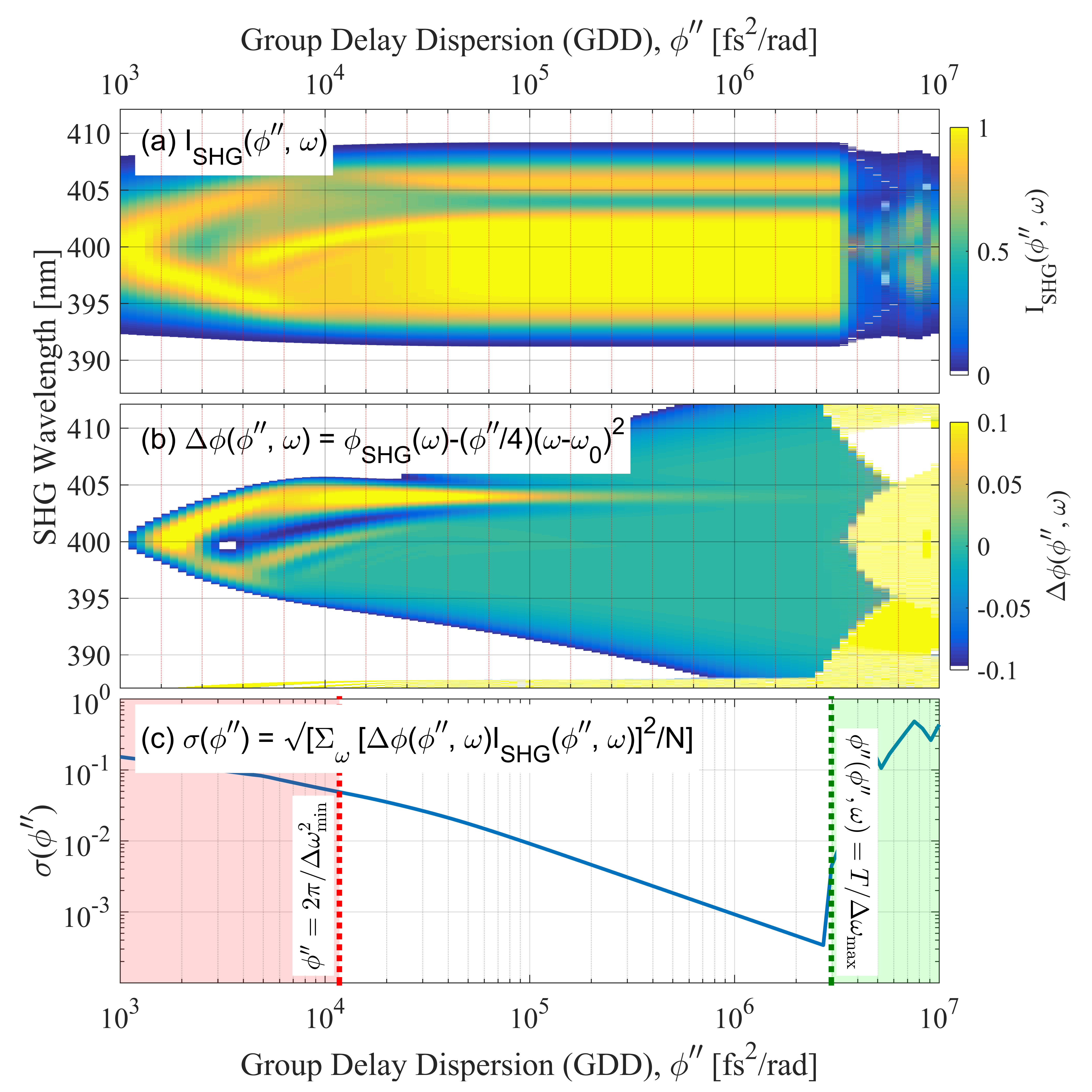}}
\caption{Dependence of the accuracy of the SHG phase approximation on GDD. (a) Simulated SHG spectrum. (b) Simulated error between actual and approximated SHG phase. (c) RMS SHG phase error with limits of validity marked (see text for details).}
\label{fig:GDD_DEP}
\end{figure}

The accuracy of the analysis described above has been verified by a comparison of  the SHG spectral phase calculated directly using \fref{eq:SFG} for $\tau=0$, and by calculating the square of the temporal electric field and is shown in \fref{fig:GDD_DEP}. The spectrum of the pulse was chosen to be a super Gaussian of bandwidth $\Delta\omega \sim 30$\,nm centered at 800\,nm with a $\Delta\omega_\text{min} \sim 3$\,nm spectral hole centred at 808\,nm and 30\% intensity. The corresponding Fourier transform limited (FTL) full width at half maximum (FWHM) pulse duration is $\Delta t \sim 55$\,fs. A linear chirp, quantified by $\phi^{\prime\prime}(\omega) = \phi^{\prime\prime}$ is applied to the spectrum and the SHG field calculated using $E_2(2\omega) \propto \mathfrak{F}^{-1}\left\{\mathfrak{F}\left[E_1(\omega)\right]^2\right\}$, where $\mathfrak{F}(\ldots)$ represents the Fourier transform. Setting $\tau=0$ in \fref{eq:SFG}, one finds that the second harmonic GDD is half the fundamental GDD, i.e. $\phi^{\prime\prime}_\text{SHG} = \phi^{\prime\prime}/2$, in the case of linear dispersion. The spectral amplitude is evenly sampled with $N=2^{14}$ points and bandwidth of $B \sim 0.3$\,rad/fs ($\Delta\lambda=100$\,nm) resulting in a temporal window of $T \sim 350$\,ps. The fine structure (i.e. spectral hole) and finite bandwidth requires that the GDD must be sufficiently large so that the chirped pulse is significantly stretched to ensure only a single quasi-monochromatic frequency is present at any given group delay, according to the criterion in \fref{eq:SPA_CRITERION} and marked by the dotted red line in \fref{fig:GDD_DEP}(c). For GDD magnitudes below this criterion, multiple frequencies interact, resulting in a structured SHG spectrum and phase. For larger GDD magnitudes, the actual SHG phase asymptotically approaches that calculated using \fref{eq:SFG_APPROX}. Note that the criterion must use the smallest significant spectral feature, which is the width of the spectral hole in this scenario, although this only provides a ``rule of thumb''. For smooth spectra, the full width at half maximum bandwidth will suffice. If the magnitude of the GDD is too large, then the pulse is stretched outside the temporal window given by the sampling rate, and is indicated by the dotted green line in \fref{fig:GDD_DEP}(c).

\section{Experiment}
\label{sec:EXPERIMENT}

\begin{figure}[htbp]
\centering
\fbox{\includegraphics[width=\linewidth]{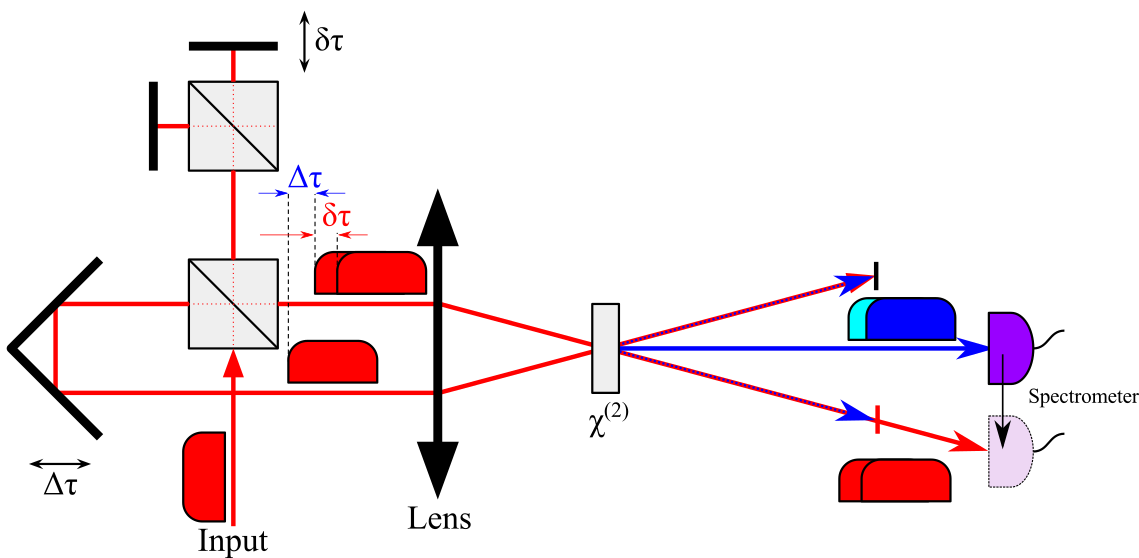}}
\caption{Schematic of experimental setup used to test the CHIMP method. An input CTP is split into two parallel paths using a beamsplitter and focused into a 200\,\textmu{}m thick type I BBO crystal. A Michelson interferometer is inserted into one of the arms before the lens to generated two collinear pulse replicas with a fixed delay $\delta\tau$. The SFG signal is spatially filtered and measured on a single array spectrometer. The SFG interferogram is measured as a function of the relative delay $\Delta\tau$ between the two non-collinear arms. The delay $\delta\tau$ is calibrated by moving the spectrometer to measure the interference between the two collinear pulses after a long pass filter.}
\label{fig:SETUP}
\end{figure}

\begin{figure}[htbp]
\centering
\fbox{\includegraphics[width=\linewidth]{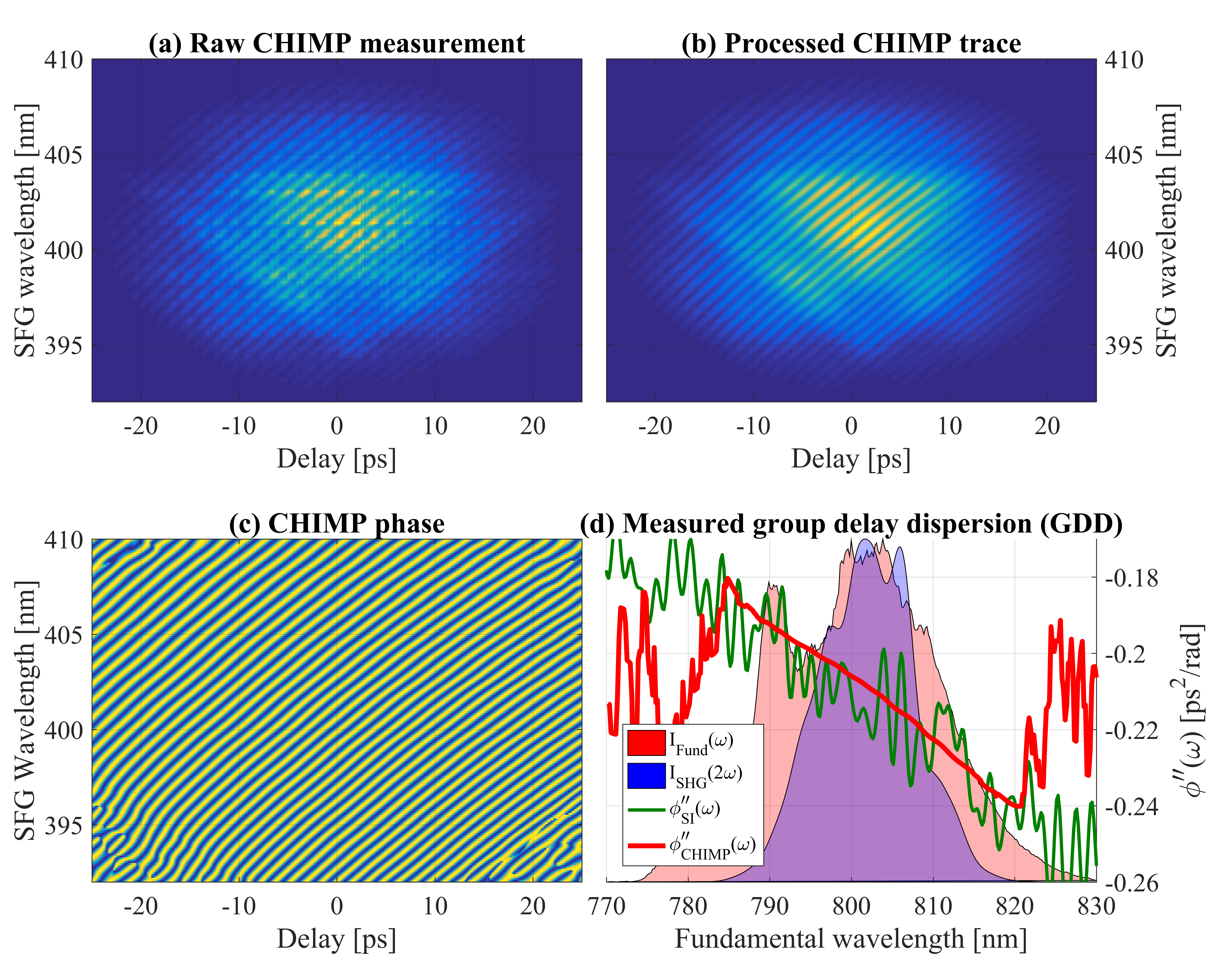}}
\caption{Experimental CHIMP results. (a) Raw measured CHIMP interferograms as a function of the delay, $\Delta\tau$ between CTP$_1$ and CTP$_{2,3}$. (b) Processed CHIMP interferogram after bandpass filtering in the Fourier domain. (c) Extracted CHIMP phase. (d) Reconstructed CTP GDD (red) using \fref{eq:CHIMP_PHASE_MODIFIED}, and independently estimated GDD from SI measurements. Shaded curves indicate fundamental spectrum (red) and SHG spectrum (blue).}
\label{fig:RESULTS}
\end{figure}

The method was experimentally verified using the setup depicted in \fref{fig:SETUP}. The delay, $\delta\tau$ between the two collinear pulses was calibrated by moving the spectrometer to measure the interference pattern of the fundamental pulses after passing a long-pass filter, and then remained fixed throughout the rest of the measurements. The delay, $\Delta\tau$, was varied using an encoded translation stage (Newport CONEX-AG-LS25-27P) and the SFG interference pattern recorded for each delay. The resulting CHIMP interferograms are plotted in \fref{fig:RESULTS}(a). A two-dimensional Fourier transform was applied to the measured trace, follwed by multiplication with a bandpass filter centred around one of the sidebands, and finally an inverse two-dimensional Fourier transform applied. The phase of this term, corresponding to the CHIMP phase described in \fref{eq:CHIMP_PHASE} is plotted in \fref{fig:RESULTS}(c). The same procedure but using a bandpass filter centred around the DC component was used and combined with twice the real part of the filtered sideband to produce the interferograms shown in \fref{fig:RESULTS}(b).

The reconstruction of the GDD proceeds slightly differently to that described previously. Due to the setup employed, there is an imbalance in dispersion between the two collinear CTPs and the single CTP. A similar analysis to that descibed above using the relationship $\phi_{2,3}^{\prime\prime}(\omega) = \phi^{\prime\prime}_1(\omega) + \Delta\phi^{\prime\prime}(\omega)$ yields a CHIMP phase of
\begin{align}
    \label{eq:CHIMP_PHASE_MODIFIED}
    \theta(2\omega, \delta\tau, \tau_{12}) &= \frac{2\tau_{12}\delta\tau + 2\Delta\phi^{\prime\prime}(\omega)\delta\tau - \delta\tau^2}{4\phi^{\prime\prime}_1(\omega) + 2\Delta\phi^{\prime\prime}(\omega)} + \omega\delta\tau\\\nonumber
    &= \alpha(\omega)\tau_{12} + \beta(\omega),
\end{align}
where $\alpha(\omega) = \delta\tau/\left[2\phi^{\prime\prime}_1(\omega) + \Delta\phi^{\prime\prime}(\omega)\right]$. Since there are many values of $\tau_{12}$, the GDD of the CTP is overdetermined, a weighted linear fit over $\tau_{12}$ using the sideband amplitude as a weighting factor can be used to determine the spectrally dependent coefficient $\alpha(\omega)$. Rearranging this and subtracting the known dispersion of the beamsplitter yeilds the desired GDD, which is plotted in \fref{fig:RESULTS}(d). To verify the accuracy of the GDD, an independent estimate of the GDD was obtained using spectral interferometry (SI) between the input and output pulses of the stretcher used before the laser amplifier. The stretcher input is assumed to be near FTL and an estimate of the GDD of the CPA was then added to the GDD of the stretcher as measured using SI. Also plotted is the SHG spectrum, obtained from the amplitude of the extracted sideband at zero delay, and an independently measured fundamental spectrum. Due to the delay between the mixing fields, not all of the spectrum is upconverted at any one time, and therefore the SHG spectrum is not as broad as the fundamental spectrum. For large magnitudes of $\tau_{12}$, the interferometric phase is very large, compensating for the low signal intensity. At small values of $\tau_{12}$, the interferometric phase is small, but the intensity is large. By performing a weighted linear fit, the precision of the GDD extracted using CHIMP far exceeds that measured using SI.

\section{Conclusions}
\label{sec:CONCLUSIONS}

It has been shown that the GDD of highly chirped broadband pulses can be measured using a simple interferometric setup and doubling crystal. The GDD is retrieved from the phase of the interferograms using a simple direct (i.e. non-iterative) reconstruction algorithm. The information can be extracted from the measured data using a Fourier filtering routine, which has the advantage of removing a significant amount of noise. This can be seen that noting the precision of the CHIMP measurements is significantly better than that of the SI measurements (smooth CHIMP curve versus the highly modulated SI curve), which is a combination of the Fourier filtering and weighted fit used in the CHIMP reconstruction. The geometry presented here was chosen for its flexibility and simplicity at demonstrating the concept of the method, but a different geometry that should enable single-shot measurements is already under investigation. It is expected that this method will prove useful in measuring the exact GDD of the various pulses in our CC-OPCA system so that we can obtain fully compressible pulses with the maximum bandwidth available.

\paragraph{Funding.} This work was supported by STFC.

\paragraph{Acknowledgements.} The authors would like to thank E. Springate for her comments on the paper.



\begin{thebibliography}{10}
\newcommand{\enquote}[1]{``#1''}

\bibitem{assion1998control}
A.~Assion, T.~Baumert, M.~Bergt, T.~Brixner, B.~Kiefer, V.~Seyfried,
  M.~Strehle, and G.~Gerber, Science \textbf{282}, 919 (1998).

\bibitem{levis2001selective}
R.~J. Levis, G.~M. Menkir, and H.~Rabitz, Science \textbf{292}, 709 (2001).

\bibitem{strickland1985compression}
D.~Strickland and G.~Mourou, Optics communications \textbf{56}, 219 (1985).

\bibitem{backus1998high}
S.~Backus, C.~G. Durfee~III, M.~M. Murnane, and H.~C. Kapteyn, Review of
  scientific instruments \textbf{69}, 1207 (1998).

\bibitem{ross1997prospects}
I.~Ross, P.~Matousek, M.~Towrie, A.~Langley, and J.~Collier, Optics
  Communications \textbf{144}, 125 (1997).

\bibitem{cerullo2003ultrafast}
G.~Cerullo and S.~De~Silvestri, Review of scientific instruments \textbf{74}, 1
  (2003).

\bibitem{butkus2004progress}
R.~Butkus, R.~Danielius, A.~Dubietis, A.~Piskarskas, and A.~Stabinis, Applied
  Physics B \textbf{79}, 693 (2004).

\bibitem{herrmann2009generation}
D.~Herrmann, L.~Veisz, R.~Tautz, F.~Tavella, K.~Schmid, V.~Pervak, and
  F.~Krausz, Optics letters \textbf{34}, 2459 (2009).

\bibitem{tavella2007dispersion}
F.~Tavella, Y.~Nomura, L.~Veisz, V.~Pervak, A.~Marcinkevi{\v{c}}ius, and
  F.~Krausz, Optics letters \textbf{32}, 2227 (2007).

\bibitem{mikhailova2011ultra}
J.~M. Mikhailova, A.~Buck, A.~Borot, K.~Schmid, C.~Sears, G.~D. Tsakiris,
  F.~Krausz, and L.~Veisz, Optics letters \textbf{36}, 3145 (2011).

\bibitem{klingebiel2013picosecond}
S.~Klingebiel, \enquote{Picosecond pump dispersion management and jitter
  stabilization in a petawatt-scale few-cycle opcpa system,} Ph.D. thesis, lmu
  (2013).

\bibitem{skrobol2012broadband}
C.~Skrobol, I.~Ahmad, S.~Klingebiel, C.~Wandt, S.~A. Trushin, Z.~Major,
  F.~Krausz, and S.~Karsch, Optics express \textbf{20}, 4619 (2012).

\bibitem{rigaud2016supercontinuum}
P.~Rigaud, A.~Van~de Walle, M.~Hanna, N.~Forget, F.~Guichard, Y.~Zaouter,
  K.~Guesmi, F.~Druon, and P.~Georges, Optics Express \textbf{24}, 26494
  (2016).

\bibitem{tang2008optical}
Y.~Tang, I.~Ross, C.~Hernandez-Gomez, G.~New, I.~Musgrave, O.~Chekhlov,
  P.~Matousek, and J.~Collier, Optics letters \textbf{33}, 2386 (2008).

\bibitem{wyatt2015ultra}
A.~S. Wyatt, P.~Oliveira, A.~Boyle, Y.~Tang, M.~Galimberti, I.~N. Ross, I.~O.
  Musgrave, C.~Hernandez, and J.~Collier, \enquote{Ultra-broadband spectral
  phase control in the vulcan 20pw upgrade front end,} in \enquote{The European
  Conference on Lasers and Electro-Optics,}  (Optical Society of America,
  2015), p. CG\_P\_1.

\bibitem{vozzi2007millijoule}
C.~Vozzi, F.~Calegari, E.~Benedetti, S.~Gasilov, G.~Sansone, G.~Cerullo,
  M.~Nisoli, S.~De~Silvestri, and S.~Stagira, Optics letters \textbf{32}, 2957
  (2007).

\bibitem{brida2009few}
D.~Brida, C.~Manzoni, G.~Cirmi, M.~Marangoni, S.~Bonora, P.~Villoresi,
  S.~De~Silvestri, and G.~Cerullo, Journal of Optics \textbf{12}, 013001
  (2009).

\bibitem{thai2011sub}
A.~Thai, M.~Hemmer, P.~Bates, O.~Chalus, and J.~Biegert, Optics letters
  \textbf{36}, 3918 (2011).

\bibitem{osvay1999efficient}
K.~Osvay and I.~N. Ross, Optics communications \textbf{166}, 113 (1999).

\bibitem{goda2013dispersive}
K.~Goda and B.~Jalali, Nature Photonics \textbf{7}, 102 (2013).

\bibitem{boivin1997206}
L.~Boivin, M.~Nuss, W.~Knox, and J.~Stark, Electronics Letters \textbf{33}, 827
  (1997).

\bibitem{mahjoubfar2015design}
A.~Mahjoubfar, C.~L. Chen, and B.~Jalali, Scientific reports \textbf{5} (2015).

\bibitem{walmsley2009characterization}
I.~A. Walmsley and C.~Dorrer, Advances in Optics and Photonics \textbf{1}, 308
  (2009).

\bibitem{frank2012invited}
F.~Frank, C.~Arrell, T.~Witting, W.~Okell, J.~McKenna, J.~Robinson, C.~Haworth,
  D.~Austin, H.~Teng, I.~Walmsley \emph{et~al.}, Review of Scientific
  Instruments \textbf{83}, 071101 (2012).

\bibitem{kim2013petahertz}
K.~T. Kim, C.~Zhang, A.~D. Shiner, B.~E. Schmidt, F.~L{\'e}gar{\'e},
  D.~Villeneuve, and P.~Corkum, Nature Photonics \textbf{7}, 958 (2013).

\bibitem{wyatt2016attosecond}
A.~S. Wyatt, T.~Witting, A.~Schiavi, D.~Fabris, P.~Matia-Hernando, I.~A.
  Walmsley, J.~P. Marangos, and J.~W. Tisch, Optica \textbf{3}, 303 (2016).

\bibitem{iaconis1998spectral}
C.~Iaconis and I.~A. Walmsley, Optics letters \textbf{23}, 792 (1998).

\bibitem{trebino1997measuring}
R.~Trebino, K.~W. DeLong, D.~N. Fittinghoff, J.~N. Sweetser, M.~A.
  Krumb{\"u}gel, B.~A. Richman, and D.~J. Kane, Review of Scientific
  Instruments \textbf{68}, 3277 (1997).

\bibitem{lozovoy2004multiphoton}
V.~V. Lozovoy, I.~Pastirk, and M.~Dantus, Optics letters \textbf{29}, 775
  (2004).

\bibitem{miranda2012simultaneous}
M.~Miranda, T.~Fordell, C.~Arnold, A.~L’Huillier, and H.~Crespo, Optics
  express \textbf{20}, 688 (2012).

\bibitem{bowlan2011complete}
P.~Bowlan and R.~Trebino, Optics express \textbf{19}, 1367 (2011).

\bibitem{wyatt2016frequency}
A.~S. Wyatt, \enquote{Frequency-resolved optical gating of highly chirped
  ultrabroadband pulses,}  (2016). In preparation.

\bibitem{takeda1982fourier}
M.~Takeda, H.~Ina, and S.~Kobayashi, JosA \textbf{72}, 156 (1982).

\end{thebibliography}

 \newcommand{\noop}[1]{}

\end{document}